\documentclass[pra,twocolumn,preprintnumbers,amsmath,amssymb,superscriptaddress]{revtex4-1}
\usepackage{dsfont}
\usepackage{color}
\usepackage{graphicx}% Incl fgfrude figure files
\usepackage{dcolumn}% Align table columns on decimal point
\usepackage{bm}% bold math
\usepackage{hyperref}
\usepackage{enumitem}
\usepackage{textcomp}
\usepackage{balance}
\usepackage{comment}
\usepackage{cleveref}
\usepackage{quantikz}
\usepackage{bbold}
\usepackage{amsthm}

\begin{document}

\title{Efficient quantum simulation of unitary gates based on atomic ensembles}

\author{Manish Chaudhary}
\thanks{manish.phys123@gmail.com}
\affiliation{Institut de Physique Nucléaire, Atomique et de Spectroscopie, CESAM, University of Liège, B-4000 Liège, Belgium}

%\author{Kaixuan Zhou}
%\affiliation{New York University Shanghai, 567 West Yangsi Road, Pudong, Shanghai 200126, China}

\author{Tim Byrnes}
\affiliation{Department of Physics, New York University, New York, NY, 10003, USA}
\affiliation{New York University Shanghai; NYU-ECNU Institute of Physics at NYU Shanghai, 567 West Yangsi Road, Pudong, Shanghai 200126, China}
\affiliation{Center for Quantum and Topological Systems (CQTS),
NYUAD Research Institute, New York University Abu Dhabi, UAE}

\date{\today}% It is always \today, today,

\begin{abstract}
We develop a technique to perform efficient unitary time evolution of a general spin Hamiltonian based on the system with identical qubits, as naturally realized by permutation symmetric atomic ensembles. We employ Cartan decomposition method to the associated Hamiltonian Lie algebra and construct the symmetry-adapted generators to implement unitary operator of the ensemble Hamiltonian. This enables the compact circuit representations of the system dynamics in terms of experimentally accessible collective spin operations and exploits the Dicke basis structure of the ensembles as an effective qudit representation. We benchmark Cartan-based evolution against both exact time evolution and the Suzuki–Trotter product formula. Our results show that the Cartan protocol accurately reproduces the operator dynamics, spin expectation values, and fluctuation observables with zero error that are substantially smaller than those obtained through Trotterization while maintaining an optimized circuit structure. In addition, scaling analysis with the ensemble size demonstrates that the Cartan error remains independent of the atom number while the corresponding Trotter error increases with the system size, highlighting the improved efficiency of the Cartan approach for a larger system size. 
%These findings establish Cartan decomposition as a promising route toward resource-efficient circuit synthesis for neutral-atom and ensemble-based quantum platforms, with direct relevance to collective simulation, spin squeezing, and many-body quantum control.
\end{abstract}

%\pacs{03.75.Dg, 37.25.+k, 03.75.Mn}% PACS, the Physics and Astronomy
\maketitle

\section{\label{sec1}Introduction}
Quantum simulation~\cite{feynman1982,lloyd1996universal} is widely regarded as one of the most promising applications of quantum information processors, as it offers the possibility of efficiently reproducing the dynamics of many-body quantum systems that are intractable on classical computers~\cite{RevModPhys.86.153,PhysRevLett.98.070201}. Finding the time evolution of a general unitary operator is an important task in quantum simulation which is used to implement the quantum circuit of a physical Hamiltonian. There are different ways to study this aspect, such as employed in imaginary time evolution~\cite{PRXQuantum.2.010342}, Suzuki–Trotterization~\cite{trotter1959product,PhysRevX.11.011020} and variational method~\cite{mcardle2019variational,cirstoiu2020variational,kalev2021quantum}. These methods are approximate in a usual sense as these involve either truncating the series expansion of the unitary operator~\cite{PhysRevLett.114.090502,ostmeyer2023optimised} or approximating the exponential term by suitable measurement operators~\cite{PhysRevLett.131.110602}. In most digital quantum simulation approaches, the target Hamiltonian is decomposed into a sum of noncommuting terms~\cite{suzuki1985decomposition}, and the corresponding time-evolution operator is approximated using Suzuki–Trotter product formulas~\cite{PhysRevX.11.011020,dhand2014stability}. While Trotterization provides a systematic route towards controllable approximations, its accuracy depends on the number of Trotter steps~\cite{ostmeyer2023optimised}, causing circuit depth and gate count to increase with simulation time and desired precision~\cite{lane2025ordering}. Consequently, Trotter errors accumulate during long-time evolution~\cite{chen2026trotter}, and the large number of gates required often exceeds the coherence times of current noisy intermediate-scale quantum (NISQ) devices~\cite{Preskill2018quantumcomputingin}.

Recent efforts have therefore focused on developing alternative Hamiltonian simulation strategies capable of reducing the circuit depth while preserving accuracy~\cite{PhysRevLett.129.070501,PhysRevA.105.032420}.
For instance, one can use the certain symmetries of the qubit Hamitonian~\cite{d2006decompositions,daugli2008general} and simplify the unitary evolution based on this symmetry Lie group~\cite{carter1995lectures}.
%For a qubit, the unitary time evolution can be implemented and optimized by employing certain symmetries of the Hamiltonian.
The method is based on Lie-algebraic methods based on Cartan decomposition~\cite{khaneja2001cartan,earp2005constructive} that separates the dynamical Lie algebra generated by a Hamiltonian into a compact subalgebra and its complementary subspace, enabling the construction of exact or near-exact factorizations of the corresponding unitary evolution~\cite{khaneja2001cartan,d2006decompositions}. In quantum computing, Cartan decompositions have played a central role in quantum circuit synthesis~\cite{PhysRevA.105.032420}, quantum compilation~\cite{alsheikh2025redcard}, and optimal gate construction~\cite{PhysRevLett.129.070501} by expressing a unitary transformation as a sequence of local operations surrounding a reduced set of nonlocal generators. %Such decompositions exploit the underlying algebraic structure of the Hamiltonian rather than relying on small-time approximations, thereby providing a fundamentally different route to quantum simulation.

The advantage of using Cartan decomposition based Hamiltonian simulation  is accounted for by the fact that it generates a fixed-depth quantum circuit that is independent of the simulation time~\cite{PhysRevLett.129.070501} in contrast to Suzuki–Trotter approach, where the circuit depth scales with the evolution time~\cite{chen2026trotter}. Moreover, such algebraic Cartan constructions generate exact simulation circuits with substantially fewer gates~\cite{PhysRevLett.129.070501,PhysRevA.105.032420} and significantly improved accuracy compared with product formula methods for broad classes of spin and fermionic models~\cite{l88j-p4f3}. %the number of required gates to implement optimized unitary evolution is minimal as compared to other approximation order . Kökcü *et al.* demonstrated that ,while simultaneously providing deeper insight into the geometric structure of the underlying dynamics. 
These developments suggest that Cartan-based approaches constitute a promising route toward resource-efficient quantum simulation on near-term and fault-tolerant quantum architectures~\cite{PhysRevA.108.052607}.

In this work, we generalize the Cartan decomposition method to the system of collective spin ensembles, where the dynamics is naturally restricted to a permutation symmetric subspace of identical atoms~\cite{PhysRevA.78.052101}. Ensemble-based implementations offer an attractive platform for quantum information processing because collective spin degrees of freedom enable an exponential reduction of the Hilbert-space dimension~\cite{moroder2012permutationally} while preserving many-body correlations and entanglement properties~\cite{PhysRevA.88.012305}. Various research works have been performed on the implementation of unitary circuits with the qubit spin Hamiltonian~\cite{PhysRevLett.129.070501,PhysRevA.105.032420,PhysRevA.108.052607}, but none of the work has been performed on permutation symmetric manifolds. By constructing the dynamical Lie algebra from collective spin operators and identifying the corresponding Cartan subalgebra, we derive compact circuit representations of the target dynamics in terms of experimentally accessible collective spin operations with an equivalent qutid picture. We further compare the resulting Cartan circuits with conventional Trotterized evolutions to estimate the reduced circuit complexity. Our results establish a systematic framework for Lie-algebraic circuit synthesis in collective quantum systems and highlight the advantages of ensemble-based quantum simulation for scalable quantum technologies.
%and demonstrate that the algebraic approach reproduces the dynamics with significantly reduced circuit complexity. 

%[TB: Maybe sell this paper as the qudit generalization of the Cartan decomposition technique] 

The paper is structured as follows.  In Sec. \ref{sec2} we review the 
Cartan decomposition method of a Lie algebra and introduce the physical system to study the time evolution of the collective spin Hamiltonian. In Sec. \ref{sec3} we develop the formalism to perform the Cartan decomposition of the Lie algebra based on collective spin operators in permutationally invariant space and introduce ensemble based gates and optimize circuit construction. In Sec. \ref{sec4} we check the numerical performance of the method by calculating spin expectation values, error with the evolution time as a function of different system sizes, and compare it with the standard Suzuki Trotter method for the spin squeezing Hamiltonian dynamics. In Sec. \ref{sec5}, we provide a possible experimental implementation to build circuit elements based on atomic ensembles. Finally in Sec. \ref{sec6}, we summarize the main conclusion of our work.

\section{Theoretical Elements} 
\label{sec2}

In this section, we first review the theory that is required to construct the Cartan decomposition method for the permutation symmetric systems. For this, we introduce the spin system and its structure that obeys a symmetry under the exchange of qubits. Then we discuss the principle behind Cartan decomposition method~\cite{d2006decompositions,drury2008constructive} that is used to split a spin Hamiltonian Lie algebra into two orthogonal subspaces. This provides an efficient way of building the time evolution of a unitary corresponding to the spin Hamiltonian.

\subsection{Spin Hamiltonian and permutation symmetry}
\begin{figure}[t]%
\includegraphics[width=\linewidth]{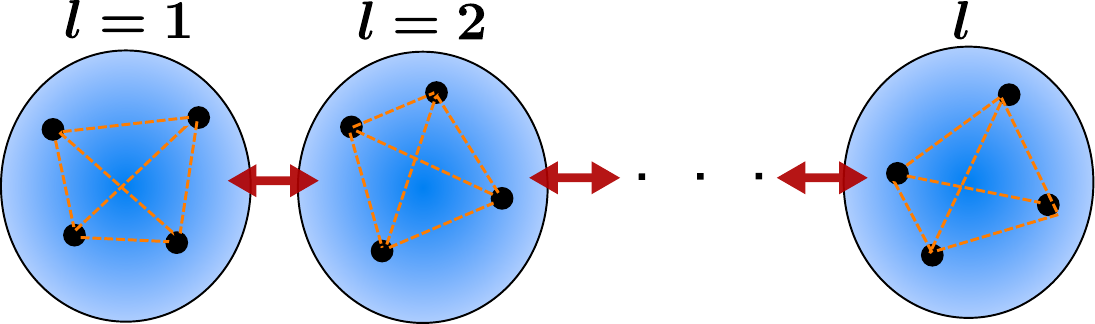}
\caption{Permutation symmetric system of $l$ interacting atomic ensembles: Each particle in an ensemble interacts with other particles in an identical way denoted by dashed lines. Neighboring ensembles are coupled through identical interactions~\eqref{ham_ensemble} denoted by arrows.}
\label{fig0}
\end{figure}

The physical system we shall consider consists of an assembly of identical spin-$1/2$ qubits such that the system is symmetric under the exchange of the particles (Fig.~\ref{fig0}) and the individual control of the particles is not necessary to consider. This is easily realized for neutral atomic ensembles, or BECs, where each atom has two populated internal states~\cite{timquantumoptics2020}. If $g_l,e_l$ represents the annihilation operators for the ground state and the excited states in the $l^{\text{th}}$ ensemble respectively, then the collective spin can be defined using the Schwinger boson operators as
\begin{align}
   S^x_l & = e^\dagger_l g_l + g^\dagger_l e_l \nonumber \\
    S^y_l & = -ie^\dagger_l g_l + i g^\dagger_l e_l \nonumber \\
     S^z_l & = e^\dagger_l e_l - g^\dagger_l g_l 
\end{align}

Since the system is symmetric under particle exchange, one defines a symmetric operator 
\begin{align}
    P & = P^{(1,2\dots n)}_{1} \otimes P^{(1,2\dots n)}_{2} \otimes \dots \nonumber \\  
    & = \prod_l P^{(1,2\dots n)}_{l}
    \label{eq:perm_symm}
\end{align}
where $P_l$ is the permutation group consisting of the set of operators that exchange the particle label in the $l^{\text{th}}$ ensemble. 

With this, the collective spin operators can also be described in terms of the sum of the individual Pauli operators for the atoms as
\begin{align}
    S^\alpha_l & =  \sum_{n=1}^N \sigma^\alpha_{l,n} 
     \label{eq:spinopsensemble}
\end{align}
where $ \alpha \in \{ x,y,z \} $ and $ \sigma^\alpha_{l,n} $ is a Pauli operator along the $ \alpha $ direction for the $n^{\text{th}}$ atom in the $l^{\text{th}}$ ensemble. %We note that $(S^\alpha_l)^2 \neq I $ in contrast to the Pauli operators $(\sigma^\alpha_{l,n})^2 = I$.
We consider the number of atoms $N$ in each ensemble to be equal for simplicity. 
The commutation relation for the spin operators follows
\begin{align}
[S^\alpha_l,S^\beta_m]=2i\epsilon_{\alpha\beta\gamma} S^\gamma \delta_{lm},   \label{commutator}
\end{align}
where $\epsilon_{\alpha\beta\gamma}$ is the Levi-Civita symbol.

Spin operators~\eqref{eq:spinopsensemble} remain invariant under the action of the symmetric operator~\eqref{eq:perm_symm}, i.e.
\begin{align}
     P S^\alpha_l P^\dagger = S^\alpha_l
\end{align}
The spin coherent states for $N$ uncorrelated atoms in an ensemble is defined as
\begin{align}
    |\theta, \phi\rangle\rangle_l & =  \frac{1}{\sqrt{N!}} \left(\cos \frac{\theta}{2} e^{-i\phi/2 }e^\dagger_l  + \sin \frac{\theta}{2} e^{i\phi/2 }g^\dagger_l \right)^N |\text{vac}\rangle \nonumber \\
    & = \sum_k c_k |k_l\rangle
    \label{spincoherentstate}
\end{align}
where $\theta, \phi$ are the angles on the Bloch sphere; $|\text{vac}\rangle$ is the vacuum state that contains no atoms and $c_k$ is the amplitude coefficient of the state when it is written in terms of the Dicke state $|k_l\rangle$ defined as,
\begin{align}
   |k_l\rangle = \frac{(e^\dagger_l)^k(g^\dagger_l)^{N-k}}{\sqrt{k!(N-k)!}} |\text{vac}\rangle
   \label{basisstate}
\end{align}
such that the Dicke states are eigenstates of the $S^z$ spin operator
\begin{align}
    S^z_l |k_l\rangle = (2k-N) |k_l\rangle.
\end{align}
It then follows that if the initial state is symmetric under particle interchange and only symmetric operators are used to manipulate the state, the state remains in the symmetric subspace \cite{timquantumoptics2020}.

%We consider the unitary evolution of the spin Hamiltonian with identical qubits with permutational symmetry. This can be defined physically with the ensemble of $N$ identical non-interacting qubits such as neutral atomic gas ensembles or Bose-Einstein condensates (BECs). 

%The system is symmetric under the exchange of the particles and one does not take into account the individual control of the particles. However, each ensemble is individually controllable.  

%We may then equivalently work using Schwinger boson operators 
%
%\begin{align}\label{eq:spinops}
   % S^x_l & =a_l^\dagger b_l+b_l^\dagger a_l \nonumber \\
   % S^y_l & =-ia_l^\dagger b_l+ib_l^\dagger a_l \nonumber \\
   %  S^z_l & =a_l^\dagger a_l-b_l^\dagger b_l  . 
%\end{align}
%
%where the bosonic annihilation operator for the qubit states are $a_l, b_l $ respectively.  Working in the completely symmetric subspace reduces the Hilbert space dimension of each ensemble from $ 2^N $ to $ N +1 $.  For convenience we shall use the bosonic formulation (\ref{eq:spinops}) henceforth.

We can construct the physical spin Hamiltonian with the possible combination of the spin operators $S^\alpha_l$ that follows the permutation symmetry~\eqref{eq:perm_symm}, the most common choice is 
\begin{align}
    \mathcal{H} = \sum_j \xi_j \mathcal{S}_j
    \label{genHam}
\end{align}
where $\mathcal{S}_j$ is the string of the collective spin operators, i.e. elements of the $l$-ensemble point group $\mathcal O_l = \{I,S^x,S^y,S^z\}^{\otimes l}$  and $\xi_j$ is the interaction coefficient. 
For instance, the spin squeezing interaction between two ensembles~\cite{kitzinger2020} is the product of two spin operators, i.e. it involves quadratic terms $\mathcal{S}_2 = S^\alpha_l S^\alpha_{l+1}$. Fig.~\ref{fig0} represents the schematic diagram of the interacting ensemble systems.

Finding the unitary $e^{-i \mathcal{H}t}$ for a particular spin Hamiltonian and its realization towards an optimized quantum circuit is an interesting problem. The usual Suzuki trotterization method can be used that involves approximating the series expansion, however,  it is less efficient in the sense that it generates the circuit depth that depends on the simulation time. We will be using the Cartan decomposition method to construct this unitary.

%BCH theorem takes into consideration all possible pairs in (\ref{eq:Ham algebra}), however Cartan decomposition simplifies by letting only few terms to be taken into consideration based on orthogonal split. Moreover, we can impose additional requirement based on exchange symmetry that further reduces the dimension of Lie algebra (\ref{eq:Ham algebra}).

\subsection{Cartan decomposition of a Lie algebra}
Here we review the formalism for decomposing the Hamiltonian algebra into two orthogonal components that was proposed in detail in Refs. \cite{carter1995lectures,d2006decompositions,drury2008constructive}. 

In order to implement optimized circuit of a particular Hamiltonian time evolution, we use its underlying Lie algebraic structure.
Let $\mathfrak{g}$ be a real semi simple real algebra associated with $\mathcal{H}$. The Cartan decomposition expresses $\mathfrak{g}$ as
\begin{equation}
\mathfrak{g} = \mathfrak{l} \oplus \mathfrak{m}
\label{cartan decomposition}
\end{equation}
where $\mathfrak{l}$ is the maximal compact subalgebra, i.e., a subspace that closes under commutation operation and it forms a Lie subalgebra and $\mathfrak{m}$ is its complementary subspace, which is not a Lie algebra and  does not close under commutation. This partition is performed by the involution function $\theta$ defined as,
\begin{equation}
    \theta (g) = -g^{T}
    \label{eq:involution}
\end{equation}
for any $g \in \mathfrak{g} $. However this choice is not unique and alternative involutions can be constructed by selecting an appropriate unitary transformation~\cite{drury2008constructive}. \newline
The action of the involution~\eqref{eq:involution} partitions the Lie algebra into two complementary subspaces, corresponding to the eigenspaces associated with the positive and negative eigenvalues, respectively with the following criterion,
\begin{align}
    \theta(\mathfrak{l}) & = \mathfrak{l} \nonumber \\
    \theta(\mathfrak{m}) & = -\mathfrak{m}
    \label{eq:orthogonalsplit}
\end{align}

\begin{figure}[t]%
\includegraphics[width=\linewidth]{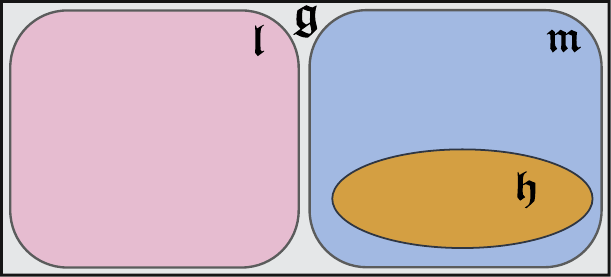}
\caption{Schematic of Cartan decomposition method: The dynamical Lie algebra $\mathfrak{g}$ is divided into subspaces $\mathfrak{l}$ and $\mathfrak{m}$, i.e. $\mathfrak{g} = \mathfrak{l} \oplus \mathfrak{m}$ based on the involution~\eqref{eq:involution}. The subspace $\mathfrak{l}$ forms a Lie sub algebra and $\mathfrak{m}$ forms an orthogonal subspace satisfying Lie bracket relations~\eqref{commutationrelations}. The maximal Abelian subalgebra $\mathfrak{h}$ is contained within $\mathfrak{m}$.}
\label{fig2}
\end{figure}

Moreover, elements of this decomposition satisfy the following Lie bracket relations,
\begin{align}
[\mathfrak{l},\mathfrak{l}] & \subseteq \mathfrak{l} \nonumber \\
[\mathfrak{l},\mathfrak{m}] & \subseteq \mathfrak{m} \nonumber \\
[\mathfrak{m},\mathfrak{m}] & \subseteq \mathfrak{l} 
\label{commutationrelations}
\end{align}
A set of mutually commuting elements can be selected from $\mathfrak{m}$ that forms a maximal abelian subalgebra such that $\mathfrak{h} \subseteq \mathfrak{m}$. The resulting Cartan decomposition of the Lie algbera $\mathfrak{g}$ and the new subspaces is shown schematically in Fig. \ref{fig2}.
%We can form a set with elements in  that are mutually commutative with each other, this set $\mathfrak{h}$ is said to form a maximal abelian algebra within $\mathfrak{m}$ such that $\mathfrak{h} \subseteq \mathfrak{m} $. The idea for Cartan split is shown 

With this construction, the Cartan decomposition theorem~\cite{carter1995lectures,d2006decompositions} confirms that there exists an element
\(K\in \mathfrak{l}\) and a Cartan element
\(h\in\mathfrak{h}\) such that an element $m \in \mathfrak{m}$ can be expressed as, 
\begin{equation}
    m = K h K^{\dagger}
\end{equation}
In practice, the unitary transformation \(K\) is constructed from
the generators \(k_j\in\mathfrak{l}\),
\begin{align}
    K(\boldsymbol{\eta}) = \prod_j e^{i\eta_j k_j},
\end{align}
where the coefficients \(\boldsymbol{\eta}=\{\eta_j\}\) are determined variationally by maximizing the Hilbert--Schmidt overlap between the rotated Hamiltonian and the chosen Cartan subalgebra, i.e.,
\begin{align}
    \boldsymbol{\eta}_\text{opt} =
   \max_{\boldsymbol{\eta}} \mathrm{Tr} \left[h K^\dagger(\boldsymbol{\eta})
    \mathcal{H} K(\boldsymbol{\eta})
    \right].
    \label{eq:trace_optimization}
\end{align}
The resulting optimal transformation identifies the Cartan frame in which the Hamiltonian is maximally aligned with the given Hamiltonian.

Choosing the physical Hamiltonian \(\mathcal{H}\in\mathfrak{m}\),
one obtains
\begin{align}
    \mathcal{H}^c = K h K^\dagger ,
    \label{cartandecom_ham}
\end{align}
which yields the Cartan time-evolution operator for the Hamiltonian as 
\begin{align}
    U(t) = e^{-i\mathcal{H}^ct} = K e^{-iht} K^\dagger .
    \label{unitary}
\end{align}
where $t$ is the dimensionless time.
Since all generators belonging to the Cartan subalgebra $\mathfrak{h}$ commute mutually, the implementation of $e^{-iht}$ reduces to a set of commuting rotations that can be
efficiently realized as a shallow quantum circuit. It is observed that the complexity of simulating the original interacting Hamiltonian is transferred to the determination of the optimal Cartan transformation $K$, while the dynamical evolution within the Cartan sector remains analytically tractable and experimentally efficient.

\section{Cartan Decomposition in the Permutationally invariant subspace}
\label{sec3}
%
%\section{Cartan Decomposition of Lie algebra of collective spin operators}
%\label{sec3}
%
In this section, we generalize the idea of Cartan decomposition method to the set of the collective spin operators and use it to simulate the Hamiltonian expressed in (\ref{genHam}) for atomic ensembles obeying permutation symmetry. 

To begin with, we construct the Hamiltonian Lie algebra $\mathfrak{g}_\text{ens}$ associated with the ensemble Hamiltonian~\eqref{genHam} by recursively evaluating all possible commutators among the Hamiltonian terms and the operators subsequently generated from them. The procedure is continued until \textit{algebraic closure} is reached, resulting in the full Hamiltonian Lie algebra as
\begin{align}
    \mathfrak{g}_\text{ens} = \text{span}\{\mathcal{S}_j,[\mathcal{S}_j,\mathcal{S}_{j'}],[\mathcal{S}_j,[\mathcal{S}_{j'},\mathcal{S}_{j''}]],\dots\}
    \label{eq:Ham algebra}
\end{align}
The resulting Lie algebra~\eqref{eq:Ham algebra} encapsulates the symmetry structure and controllable dynamics of the system and serves as the starting point for the Cartan decomposition of the associated unitary evolution.

%We provide a general theory to perform Cartan decomposition with this structure, and in the next section, we work out specific examples.

Permutation symmetry plays a crucial role in reducing the size of the dynamical Lie algebra~\cite{PhysRevA.78.052101}, thereby simplifying the resulting circuit construction based on collective spin degrees of freedom. The key simplification arises from the fact that symmetry decomposes the full Hilbert space into irreducible collective-spin sectors~\cite{thompson2008angular}. Specifically, under the joint action of $\mathfrak{su}(2)$ and the permutation group~\eqref{eq:perm_symm}, one obtains
\begin{align}
    \mathcal{H} = \underset{s}{\oplus} (\mathcal{H}_s \otimes \mathcal{M}_s)
\end{align}
where the total spin $s$ assumes values $0,\frac{1}{2}, 1, \dots, \frac{N}{2}$ arising from the allowed addition rules for the spin angular momentum. The part $\mathcal{H}_s$ carries an irreducible spin-$s$ representation with dimension $2s+1$, while the part $\mathcal{M}_s$ is a multiplicity space carrying permutation symmetry~\cite{moroder2012permutationally}. 

The symmetric part of the Hamiltonian corresponds to the space with Dicke basis in the maximal spin length $s=\frac{N}{2}$ with unit multiplicity such that 
\begin{align}
  \mathcal{H} =  P \mathcal{H} P^\dagger
  \label{Ham_invariance}
\end{align}
satisfying 
\begin{align}
    [\mathcal{H} , P] = 0
\end{align}
With this, the full symmetric operator algebra~\eqref{eq:Ham algebra} consists of an irreducible tensor operator of rank $q$ under global $\mathfrak{su}(2)$ rotation as,
\begin{align}
   \mathfrak{g}_\text{ens}= \overset{N}{\underset{q=0}{\oplus}} \mathcal{T}^{(q)}
   \label{permutation_liealgebra}
\end{align}
each sector with dimension $\text{dim}(\mathcal{T}^{(q)})=2q+1$, where $\mathcal T^{(0)}$ corresponds to the scalar sector comprising rotation invariant spin terms of the form $\vec S_l.\vec S_m$, $\mathcal T^{(1)}$ to the collective-spin generators~\eqref{eq:spinopsensemble}, and $\mathcal T^{(2)}$ to bilinear product of the spin operators with terms $S^\alpha_l S^\beta_m$, and similarly the higher-rank sectors describe increasingly nonlinear collective correlations.

Symmetry ensures that any commutator algebra of this tensor structure is always closed and lies in the symmetric space, i.e.,
\begin{align}
    [\mathcal{T}^{(q_1)}, \mathcal{T}^{(q_2)}] \subseteq \overset{q_1+q_2}{\underset{q'=|q_1-q_2|} {\oplus}} \mathcal{T}^{(q')}
    \label{eq:alg_closure}
\end{align}
that is identical to the familiar $\mathfrak{su}(2)$ algebraic structure~\eqref{commutator}.

This is in contrast to calculating the required spin Hamiltonian algebra~\eqref{eq:Ham algebra} for a system of distinct spins ($\propto  \sum_{n=1}^N \sigma^\alpha_{l,n} $) with increased algebraic complexity, which, in general, increases rapidly with $N$ as shown in the Appendix~\ref{appendix_1}. The lack of permutation symmetry leads to a much larger operator space and a more involved Cartan decomposition procedure. For instance, for a generic local controls and interactions the associated dynamical Lie algebra can grow to $\mathfrak{su}(2^N)$ with dimension $4^N-1$. In contrast, for permutation-symmetric collective-spin systems, the dynamics can be restricted to the symmetric subspace, whose Hilbert-space dimension scales only as O$(N+1)$, providing a substantially more compact representation of the system dynamics (Appendix~\ref{appendix_1}).

%For comparison, in Appendix, we discuss the increased algebraic complexity associated with the systems of distinguishable particles, where such symmetry is absent that 

%\textcolor{red}{(It appears that the algebra defined by (\ref{ensemble_hamiltonian algebra}) is countably infinite as it appears with different powers of the spin operator that has to be truncated. Some thoughts on this... maybe dealing with spin-1 Hamiltonian as it appears that $(S^j)^n =S^j$ for odd values of $n$.)}

\subsection{Example: Two ensemble based interaction} 
We consider the two-axis spin-squeezing interaction between two atomic ensembles, which has previously been investigated to generate spin-squeezed entangled states~\cite{kitzinger2020}. Here, we employ the Cartan decomposition to find exact Lie-algebraic structure of its dynamics, time evolution operator and propose a fully optimized quantum circuit based on the ensemble gates. 
The squeezing Hamiltonian is defined as,
\begin{align}
    \mathcal{H}_{\text{ens}} & =  S^x_1 S^x_2 +  S^y_1 S^y_2  \nonumber \\
    & = \frac{1}{2} (S^+_1 S^-_2 + S^-_1 S^+_2) \nonumber \\
    & \in \{ \mathcal{T}^{(0)}, \mathcal{T}^{(2)} \}
    \label{ham_ensemble}
\end{align}
where $S^{\pm}_l = S^x_l \pm i S^y_l$ are the raising and lowering spin operators for $l^{\text{th}}$ ensemble.

The Hamiltonian Lie algebra $\mathfrak{g}_{\text{ens}}$ is obtained by computing the commutator of the various spin terms~\eqref{eq:Ham algebra} for the various spin sectors satisfying~\eqref{eq:alg_closure}.
At the first level, the exact expression for the commutator between $C_1$ and $C_2$ is calculated  as
\begin{align}
    [C_1,C_2] & =  \frac{i}{2} \Bigg[ M\left( J_2(J_2+1)-J_1(J_1+1) \right )+ \nonumber \\ 
   & \left(J_1(J_1+1) + J_2(J_2+1) + \frac{M^2}{2}\right)D - \frac{D^3}{2}\Bigg]  
   \label{eq:commutation_C1C2}
\end{align}
where 
\begin{align}
    C_1 & = \frac{1}{2} (S^+_1 S^-_2 + S^-_1 S^+_2) \in  \mathcal{T}^{(0)} \oplus \mathcal{T}^{(2)} \nonumber \\
    C_2 & = \frac{1}{2i} (S^+_1 S^-_2 - S^-_1 S^+_2) \in \mathcal{T}^{(1)} \nonumber \\
    D & = S^z_1 - S^z_2 \in \mathcal{T}^{(1)}
\end{align}
and $M=S^z_1+S^z_2$ is the total spin angular momentum of two ensembles. $J_1$ and $J_2$ are the spin angular momentum numbers for the first and second ensembles respectively.

It is easy to show that,
\begin{align}
    [M,C_1]= [M,C_2] = 0
\end{align}
Hence, apart from constant contributions and mutually commuting terms that do not generate new elements under commutation, the nontrivial generators of the dynamical Lie algebra ($\mathfrak{g}_{\rm ens}$) arise from the operator structures appearing in Eq.~\eqref{eq:commutation_C1C2}.
\begin{align}
   [C_1,C_2] & = a_1 D + a_2D^3 \equiv f_o(D) \nonumber \\
   & \in \mathcal{T}^{(1)} \oplus \mathcal{T}^{(3)} 
\end{align}
where $a_1,a_2$ are the constant values for a given $N$ and the operator function $f_o(D)$ represents the \textit{odd} polynomial powers in $D$.

At next level, the contribution of the first and third order term in $D$ evaluates to
\begin{align}
[C_1,D] & =  -2iC_2 \nonumber \\
    [C_2,D] & =  2iC_1  \nonumber \\
[C_1,D^3] & =  -2i(C_2 D^2 + DC_2D + D^2 C_2 ) \nonumber \\
    [C_2,D^3] & = 2i(C_1 D^2 + DC_1D + D^2 C_1 )
\end{align}
introducing new terms of the product of the polynomial powers $f(D)$ with $C_1$ and $C_2$.

Hence, the full commutation algebra $\mathfrak{g}_{\text{ens}}$ is calculated as
%\begin{align}
 % \{ C_1, C_2, f_{e,o}(D), f_{e,o}(D) C_1, f_{e,o}(D) C_2 \}
%\label{Ham_algbera_ensemble}  
%\end{align}
\begin{align}
  \{ C_1, C_2, f_{e,o}(D),  F_{e,o;1,2}(D)\}
\label{Ham_algbera_ensemble}  
\end{align}
where $f_{e,o}(D)$ are the functions of even or odd powers of $D$ respectively
and the symmetrized product between $f_{e,o}(D)$ with $C_1$ and $C_2$ is defined by
\begin{align}
   F_{e,1}(D) & =  \frac{f_{e}(D) C_{1} + C_{1} f_{e}(D)}{2} \nonumber \\
   F_{o,1}(D) & =  \frac{f_{o}(D) C_{1} + C_{1} f_{o}(D)}{2}  \nonumber\\
   F_{e,2}(D) & =  \frac{f_{e}(D) C_{2} + C_{2} f_{e}(D)}{2} \nonumber \\
   F_{o,2}(D) & =  \frac{f_{o}(D) C_{2} + C_{2} f_{o}(D)}{2} 
\end{align}

\subsection{Optimized unitary}
To construct an efficient Cartan decomposition of the ensemble Hamiltonian, we introduce the involution function as
\begin{align}
    \theta_{\text{ens}}(g) = -AgA^\dagger
    \label{inv_ensemble}
\end{align}
where $A$ denotes the swap operator between two systems $1 \leftrightarrow 2$ and $g \in \mathfrak{g}_{\text{ens}}$. The action of~\eqref{inv_ensemble} partitions the dynamical Lie algebra~\eqref{Ham_algbera_ensemble} into the Cartan pair
\begin{align}
    \mathfrak{g}_{\text{ens}} = \mathfrak{l}_{\text{ens}} \oplus \mathfrak{m}_{\text{ens}}
\end{align}
The corresponding subspaces are spanned by
\begin{align}
    \mathfrak{l}_{\text{ens}} & = \{C_2, f_o(D),  F_{e,2}(D), F_{o,1}(D)   \} \nonumber \\
    \mathfrak{m}_{\text{ens}} & = \{C_1,f_e(D), F_{e,1}(D), F_{o,2}(D)   \} 
    \label{cartan_gen_ensemble}
\end{align}
satisfying the standard symmetric-space relations~\eqref{commutationrelations}
The maximal abelian group $\mathfrak{h}_{\text{ens}}\in\mathfrak{m}_{\text{ens}}$  is chosen as
\begin{align}
   \mathfrak{h}_{\text{ens}} = \{C_1 \} 
   \label{hensemble}
\end{align}
Consequently, the ensemble Hamiltonian based on the Cartan construction can be expressed as~\eqref{cartandecom_ham}
\begin{align}
\mathcal{H}^c_{\mathrm{ens}} =
K_{\mathrm{ens}}
 h 
K_{\mathrm{ens}}^\dagger ,
\qquad
h\in\mathfrak{h}_{\mathrm{ens}}, 
\label{cartanhamiltonian}
\end{align}
where
\begin{align}
K_{\mathrm{ens}} =
\prod_j e^{i\eta_j k_j},
\qquad
\forall k_j\in\mathfrak{l}_{\mathrm{ens}}.
\label{kensemble}
\end{align}
The coefficients $\{\eta_j\}$ are determined variationally by maximizing the overlap between the rotated Hamiltonian and the Cartan subalgebra~\eqref{eq:trace_optimization}. For the present ensemble model with $N=3$, the optimization yields
\begin{align}
    (\eta_1, \eta_2,\eta_3, \eta_4) = \left(0, 0, -\frac{\pi}{11}, \frac{\pi}{7}  \right)
\end{align}

\begin{figure*}[t]%
\includegraphics[width=\linewidth]{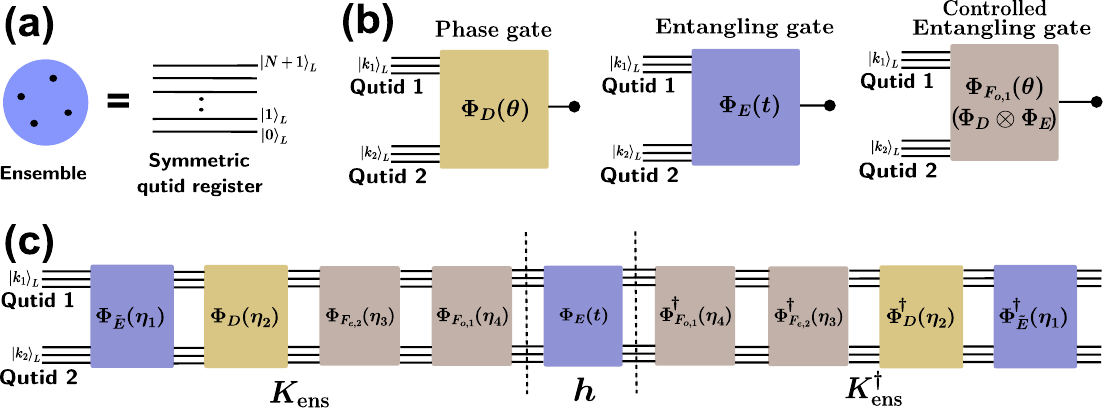}
\caption{Ensemble based optimized circuit implementation of the interaction Hamiltonian~\eqref{ham_ensemble}: (a) Ensemble as a symmetric Qutid: An ensemble of $N$ identical spins restricted to the fully symmetric Dicke manifold is represented as an effective $(N+1)$-dimensional collective qudit register. The Dicke basis states encode the collective excitation sectors of the ensemble; (b) Various ensemble gates in the collective-qudit encoding: the Dicke-state mapping~\eqref{Dicke basis} enables a circuit realization for various collective gates. We consider collective Phase gate, $\Phi_D(\theta)$, Entangling gate $\Phi_{E,\tilde{E}}(t)$ and controlled Entangling gate $\Phi_{F_{e,o;j}}(\theta)$. Action of these gates on Dicke basis is defined in Eqs.~\eqref{phasegate_ensemble},~\eqref{entanglinggate_ensemble},~\eqref{controlled_entanglinggate_ensemble} respectively; (c) Implementation of the unitary circuit~\eqref{unitary_cartan} for the collective interaction Hamiltonian~\eqref{ham_ensemble} using Cartan decomposition method where the circuit is decomposed into layers $K_\text{ens},h$ and $K^\dagger_\text{ens}$ using ensemble based gates.}
\label{fig4}
\end{figure*}

The resulting Cartan circuit realizes the time-evolution operator of the ensemble Hamiltonian in the form
\begin{align}
 U_{\text{Cartan}}(t) =  \left(\prod_j e^{i \eta_j k_j}\right) e^{-iht} \left(\prod_j e^{i \eta_j k_j}\right)^\dagger 
   \label{unitary_cartan}
\end{align}
thereby reducing the simulation of the full interacting Hamiltonian to evolution within the Abelian sector $\mathfrak{h}_{\text{ens}}$.
To check the accuracy of this construction, we compare it with the exact evolution operator defined as 
\begin{align}
    U_{\mathrm{Exact}}(t) = e^{-i\mathcal{H}_\text{ens}t}
    \label{unitary_exact}
\end{align}
as well as with Suzuki Trotterized approximation 
\begin{align}
    U_{\text{Trotter}}(t) = (e^{-iS^x_1S^x_2 \delta t}e^{-iS^y_1S^y_2 \delta t})^{n_{t}}
    \label{unitary_suzukitrotter}
\end{align}
where $\delta t = t/n_t$ and $n_t$ denotes the number of Trotter steps.
This comparison verifies the accuracy and efficiency of the Cartan-based implementation compared to conventional digital simulation protocols.

\subsection{Ensemble based circuit implementation}
Implementing unitary~\eqref{unitary_cartan} to an equivalent quantum circuit requires ensemble based gates. In this section, we provide a proposal for implementing basic circuit elements employing permutation symmetric systems.

An effective way to build this is to use symmetry arguments, as each ensemble of $N$ spin-$1/2$ particles is restricted to the fully symmetric Dicke manifold of dimension $N+1$, thus defining an effective collective qudit as shown in Fig.~\ref{fig4}(a). 
Then, collective-spin operators $S^\alpha_l$ act as logical generators on this encoded qudit space.

For instance, the collective angular momentum eigenstate $|J,M_J\rangle$ constitutes an effective basis for a logical qutid such that
\begin{align}
    S^z |J=N/2,M_J\rangle = M_J |J=N/2,M_J\rangle
\end{align}
where $M_J$ takes values between $[-J,J]$ in integral steps and represents the orientation of the total spin quantum number $J$ along the $z$ direction.

The mapping to the symmetric qudit is evident as it defines a basis
\begin{align}
    |k\rangle_L = \left|J=\frac{N}{2},M_J=k-\frac{N}{2}\right\rangle
    \label{Dicke basis}
\end{align}
with dimension $N+1$.
For example, with $N=2$, the symmetric Dicke basis 
\begin{align}
    |0\rangle_L & \equiv | J=1, M_J=-1\rangle \nonumber \\
    |1\rangle_L & \equiv | J=1, M_J=0\rangle \nonumber \\
    |2\rangle_L & \equiv | J=1, M_J=1\rangle 
\end{align}
behaves exactly like a logical qutrit. In terms of circuit terminology, an ensemble can be viewed as a \textit{physical unit} representing a logical $(N+1)$-level register (qutid).

With this notation, we can easily understand the action of the important gates as
\begin{align}
    D |k_1,k_2\rangle_L &= 2(k_1-k_2) |k_1,k_2\rangle_L \nonumber \\
   e^{-i D \theta} |k_1,k_2\rangle_L & = e^{-2i (k_1-k_2) \theta} |k_1,k_2\rangle_L
    \label{phasegate_ensemble}
\end{align}
$e^{-i D \theta}$ is a collective \textit{phase gate} that gives information about the relative phases between collective qutids. In the ensemble circuit representation, we denote this gate by $\Phi_D(\theta)$ as illustrated in Fig.~\ref{fig4}(b).

Similarly, the action of the term $S^+_1S^-_2$ is understood as transferring the population among different Dicke levels,  
\begin{align}
    S^+_1S^-_2 |k_1,k_2\rangle_L & =  |k_1+1,k_2-1\rangle_L 
\end{align}
Hence, the operator $C_{1}$ acts as,
\begin{align}
    C_{1} |k_1,k_2\rangle_L & \propto   |k_1+1,k_2-1\rangle_L +  |k_1-1,k_2+1\rangle_L 
     \label{entanglinggate_ensemble}
\end{align}
that creates a superposition of different excitation partitions generating entanglement. Therefore, time evolution $e^{-iC_1t}$ is an \textit{entangling gate} that acts on the Dicke-register space by coherently redistributing collective excitations between ensembles. This gate is denoted by $\Phi_E(t)$ as shown in the Fig.~\ref{fig4}(b).

Using the Holstein Primakoff approximation, the collective spin operators map in terms of two bosonic modes $a,b$ as
$S^+_{1} (S^-_{1})\to a^\dagger(a)$ and $S^+_{2} (S^-_{2})\to b^\dagger(b)$.
Hence, it follows that 
\begin{align}
    C_1^{\text{HP}}  & \sim   (a^\dagger b + b^\dagger a)
\end{align}
is effectively similar to a bosonic beam splitter transformation~\cite{PhysRevResearch.4.023054}.
The same interpretation follows for the action of the operator $C_2$ and we represent it schematically as $\Phi_{\tilde{E}}(t)$ in Fig.~\ref{fig4}(b). 

The action of the product operator $e^{-iF_{e,o;j}}$ represents the combined action of the collective phase gate and the entangling gate accordingly. For instance 
\begin{align}
    D C_1 |k_1,k_2\rangle_L  \propto & (k_1-k_2+2)  |k_1+1,k_2-1\rangle_L  \nonumber \\ & + (k_1-k_2-2)   |k_1-1,k_2+1\rangle_L 
    \label{controlled_entanglinggate_ensemble}
\end{align}
In a circuit picture, $e^{-i D C_1 t}$ can be viewed as a \textit{controlled entangling gate}, where the exchange strength is conditioned on the relative phase of the qutids set by the action of the operator $D$. The similar interpretation follows for the higher power products $f_{\text{e,o}}(D)$ with $C_1(C_2)$ respectively. We choose it to represent schematically as $\Phi_{F_{e,o;j}}(\theta)$ while using in the circuit in Fig.~\ref{fig4}(b). The experimental realization of these gates has been discussed in Sec.~\ref{sec5}.

With these definitions, we construct the optimized quantum circuit to realize the unitary evolution of the permutation symmetric spin Hamiltonian~\eqref{ham_ensemble} in Fig.~\ref{fig4}(c).

\section{Numerical Performance}
\label{sec4}
In this section, we evaluate the numerical performance of the optimized unitary~\eqref{unitary_cartan} constructed using the Cartan decomposition technique and calculate the time evolution of physical quantities such as the spin expectation values and variances of the conjugate spin components. We compare the time evolution of the system against the exact unitary~\eqref{unitary_exact} and the approximate unitary using Suzuki Trotter~\eqref{unitary_suzukitrotter} method.

\subsection{Average spin expectation values and variances}
\begin{figure}[t]%
\includegraphics[width=\linewidth]{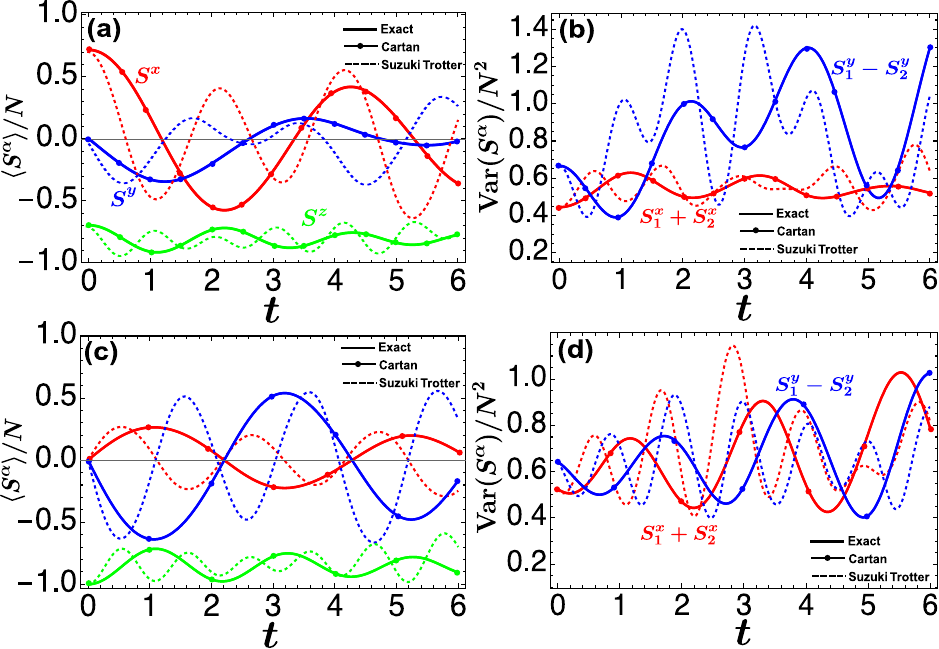}
\caption{Time evolution of system dynamics using exact and Cartan unitary: (a),(c) the collective spin expectation values~\eqref{spin_expect};  (b),(d) the variances~\eqref{spin_variances} in the combination of the spin components as a function of the interaction time $t$ for different initial states~\eqref{spincoherentstate} for $N=3$ per ensemble. Solid curves denote the exact quantum evolution, while the dashed curves correspond to the Suzuki Trotter based dynamics. The initial spin polarized state is chosen as $|\frac{\pi}{2},0 \rangle \rangle_1 \otimes |\frac{\pi}{2},0 \rangle \rangle_2$ for (a),(b), while the initial spin polarized state is chosen as $|\frac{\pi}{4},\frac{\pi}{8} \rangle \rangle_1 \otimes |\frac{\pi}{4},\frac{\pi}{8} \rangle \rangle_2$ }
\label{fig5}
\end{figure}

The average spin expectation values are expressed as,
\begin{align}
    \langle S^\alpha_l(t)\rangle = \frac{\langle \psi(t)|S^\alpha_l|\psi(t)\rangle}{\langle \psi(t)|\psi(t)\rangle}
    \label{spin_expect}
\end{align}
where the time evolved wavefunction of the system through unitary $U(t)$ is 
\begin{align}
    |\psi(t)\rangle = U(t) |\psi(0)\rangle
\end{align}
and $|\psi(0)\rangle$ is the initial wavefunction of the ensemble~\eqref{spincoherentstate} and is written in terms of the Dicke basis states~\eqref{Dicke basis}
\begin{align}
    |\psi(0)\rangle = \sum_{k_1,k_2} c_{k_1,k_2} |k_1,k_2\rangle_L
\end{align}
where $c_{k_1,k_2}$ are the amplitude coefficients of the state such that
\begin{align}
    \sum_{k_1,k_2} |c_{k_1,k_2}|^2 =1
\end{align}
Fig.~\ref{fig5}(a) calculates the time evolution of the collective spin expectation values~\eqref{spin_expect} obtained from the exact dynamics~\eqref{unitary_exact} generated by the Hamiltonian and compares with the Cartan decomposition evolution~\eqref{unitary_cartan} and Suzuki Trotter evolution~\eqref{unitary_suzukitrotter}. It is observed that the exact dynamics (solid lines) exhibit the coherent transfer of spin polarization between the two ensembles. For a given spin polarized initial state~\eqref{spincoherentstate}, $S^x_l$ spin expectation value undergoes large-amplitude oscillations, while $S^z_l$ develops a substantial negative value and $S^y_l$ remains comparatively small throughout the evolution, indicating that the dynamics occur primarily within the $x$-$z$ plane of the collective Bloch sphere. The Cartan evolution (points) matches the qualitative structure of the exact time evolution, implying that the chosen Cartan generators~\eqref{cartan_gen_ensemble} give rise to exact dynamics~\eqref{unitary}. On the other hand, the approximate Suzuki unitary~\eqref{unitary_suzukitrotter} shows large deviations in the time evolved spin expectation values, which implies that it does not resemble with the exact system dynamics. For different initial spin polarized state (Fig.~\ref{fig5}(c)), spin dynamics from exact and Cartan unitary are in agreement with each other while Suzuki Trotter shows deviation.

We also plot the variances of a certain combination of spin components as the dynamics generate spin squeezed states as studied in~\cite{kitzinger2020}.
We define the variance in the uncertainty for a spin observable $S^\alpha_l$ as
\begin{align}
    \text{Var}(S^\alpha_l) = \langle (S^{\alpha}_{l})^2\rangle - \langle S^\alpha_l\rangle^2
    \label{spin_variances}
\end{align}
In Fig.~\ref{fig5}(b) it is clearly seen that the variance Var($S^y_1+S^y_2$) is reduced below its initial level and the variance in the conjugate direction is increased Var($S^x_1-S^x_2$) signaling the emergence of spin squeezing as the exact  time evolution redistributes the quantum noise between conjugate collective modes. The Cartan based evolution captures a similar trend in squeezing dynamics in agreement with the exact time evolution. In particular, it matches the timing of the squeezing minima and the corresponding anti-squeezing maxima. However, the Suzuki Trotterization fails to capture these features.
This shows the advantage of Cartan-decomposition-based simulation as it can compress the dynamics into a shallow circuit while still preserving the principal many-body structure of the exact evolution which is true for any initial spin state (Fig.~\ref{fig5}(d)).

\subsection{Error}
We estimate the closeness of the unitary matrix of the Cartan or Suzuki trotter method with the exact unitary evolution matrix as the norm measure
\begin{align}
    \varepsilon_{\text{Cartan,Trotter}}(t) = \frac{|| U_{\text{Exact}}(t)- U_{\text{Cartan,Trotter}}(t)||}{N}
    \label{error}
\end{align}
where $||A||=\sqrt{Tr(A^\dagger A)}$ is the norm value for a matrix $A$. This can also be interpreted as the error measure, as it measures the distance between the approximate evolution operator and the exact unitary evolution generated by the exchange Hamiltonian.

\begin{figure}[t]%
\includegraphics[width=\linewidth]{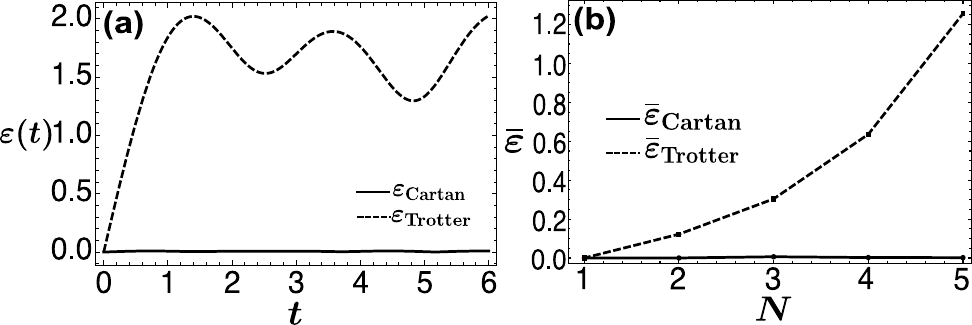}
\caption{Performance of the optimized Cartan unitary: (a) the error~\eqref{error} between the exact unitary evolution and the Cartan (Solid curve) and approximate Suzuki Trotter unitaries (Dashed curve) as a function of the interaction time $t$ for $N=3$ per ensemble;  (b) Scaling of the time-averaged error~\eqref{eq:timeaverged error} with the ensemble size $N$ for Cartan-decomposed and Suzuki–Trotter quantum circuits.}
\label{fig6}
\end{figure}

Fig.~\ref{fig6}(a) shows the error value for the Cartan-decomposed evolution and the Suzuki–Trotter approximation. It can be seen that the Cartan error remains zero as it matches the exact time unitary operation and does not involve any approximation.
\begin{align}
    \varepsilon_{\text{Cartan}} = 0
\end{align}
Also, the Cartan error is less than the error in Trotter method throughout the evolution. 
\begin{align}
    \varepsilon_{\text{Cartan}} < \varepsilon_{\text{Trotter}}
\end{align}
The Trotter error grows rapidly  and remains substantially larger, which is accounted for by the fact that it approximates the evolution by repeated products of noncommuting pieces, causing local truncation errors to accumulate during propagation~\cite{chen2026trotter}. We derive the error bound for the Suzuki–Trotter decomposition in Appendix~\ref{appendix_2}. On the other hand,  Cartan based circuit construction preserves the underlying Lie algebraic structure of the Hamiltonian dynamics more faithfully than the product-formula decomposition. This offers a significantly more accurate representation of the system dynamics than the corresponding Trotterized evolution for the same computational complexity. We calculate Cartan error bound arising from the parameter imperfections in Appendix~\ref{appendix_2}.

\subsection{Scaling with Ensemble size}
We study the dependence of the error averaged over the simulation time as a function of the ensemble size.
The averaged error is defined as,
\begin{align}
    \bar{\varepsilon} =\frac{1}{T}\sum_{j} \varepsilon(t_j)
    \label{eq:timeaverged error}
\end{align}
where $T$ is the total simulation time and the averaging is performed over the sampled evolution times.

Fig.~\ref{fig6}(b) compares the mean error of the Cartan method with that of Suzuki Trotter method as the ensemble size increases. As we have seen the Cartan method generates exact time dynamics for the system, hence it's mean value is also zero and is independent of the ensemble size. 
However, mean error for the Suzuki trotter method has a non zero value and is larger than that of Cartan method. One can see that for small system size, it is relatively low error, however, as the system size increases, the mean error of the Trotter approximation grows rapidly. We study the dependence of the errors on the system size in details in Appendix~\ref{appendix_2}.

This demonstrates that the Cartan generators capture the dominant dynamical directions of the collective-spin manifold more efficiently than a fixed-order product-formula expansion.
The better scaling of the Cartan approach originates from its Lie-algebraic construction. Rather than approximating the dynamics through repeated short-time decompositions, as in Suzuki–Trotter methods, the Cartan circuit incorporates collective generators that directly encode the symmetry-adapted structure of the permutation-invariant spin ensemble. 
Consequently, the required circuit depth remains comparatively compact, while  the essential many-body correlations responsible for collective dynamics and spin squeezing are preserved.

\section{Experimental Implementation}
\label{sec5}
In this section, we discuss the experimental viability of the physical system and ensemble gate operations in the optimized circuit construction described previously.  A key advantage of the present approach is that it does not require microscopic control or individual addressing of the constituent qubits in an ensemble. Consequently, the protocol is naturally suited to physical platforms composed of permutation-symmetric spin ensembles.
Such system involves the use of hyperfine states of $^{87}$Rb atoms, specifically $F=1,M_F=-1$ and $F=2,M_F=1$, which forms an effective two-level atom system and has been widely used for quantum information processing and quantum metrology~\cite{pezze2018quantum}. Such a system of collection of indistinguishable atoms addressing permutation symmetry can be implemented using either hot or cold atomic ensembles or BECs~\cite{timquantumoptics2020}.  

The circuit construction to implement unitary terms requires only a small set of experimentally accessible operations, namely collective spin rotations, ensemble phase gates~\eqref{phasegate_ensemble}, entangling gates between two ensembles~\eqref{entanglinggate_ensemble}, and controlled entangling operations~\eqref{controlled_entanglinggate_ensemble}. 
Quantum non demolition (QND) measurements is an effective method to determine the relative phases of ensembles without introducing backaction effects~\cite{PhysRevLett.112.233602,kuzmich1998atomic} that shows the feasibility of realizing the ensemble based phase gates~\eqref{phasegate_ensemble}.
%The only operations required consist of collective spin rotations, relative phase measurements~\eqref{phasegate_ensemble} of the ensemble, and entangling gate~\eqref{entanglinggate_ensemble} between two ensembles. 
Furthermore, QND interactions have been widely employed as an effective method for generating spin squeezing and inducing entanglement generation in atomic ensembles~\cite{julsgaard2001experimental,aristizabal2021quantum,PhysRevA.105.022443,PhysRevA.108.032420}, providing a direct experimental route for realizing the entangling gates required in the optimized Cartan circuit.
In addition, controlled entangling operations~\eqref{controlled_entanglinggate_ensemble} can be employed by combining collective phase shifts with these entangling interactions, allowing the complete gate set required in the circuit construction.

Ideally, the Cartan decomposition exactly reproduces the target unitary evolution within the chosen Lie algebra yielding zero error, however imperfections arising from ensemble gates implementation such as finite measurement efficiency, atom loss, photon loss and spontaneous emission may introduce deviations from the ideal dynamics.
Nevertheless, extensive theoretical and experimental studies have demonstrated that QND based collective operations remain remarkably robust against these dominant decoherence mechanisms, as noted in Refs.~\cite{PhysRevLett.122.163601,gao2023optical,PhysRevA.97.063613,PhysRevLett.111.253001,gao2025photon}.  Overall, we estimate that the proposed framework provides a realistic route towards implementing optimized quantum circuits for collective spin Hamiltonians on existing neutral-atom and atomic-ensemble architectures.

%\section{Experimental implementation}
%\label{sec5}

\section{Summary and Conclusions}
\label{sec6}
In this work, we introduce a Cartan decomposition framework to implement the effective time unitary of the physical Hamiltonian  based on permutation-symmetric spin systems and collective atomic ensembles. By exploiting the reduced-dimensional symmetric subspace, we construct a hierarchy of collective generators that enables the decomposition of ensemble Hamiltonian into experimentally accessible unitary blocks. Unlike conventional Suzuki Trotter method that factorizes the product terms, the present approach takes advantage of the collective spin symmetries to optimize the circuit representations directly in the Dicke manifold. This translates the many-body collective dynamics into efficient quantum circuits while preserving the underlying permutation symmetry associated with the physical system.

We consider an exchange spin squeezing interaction as a representative model and analyze Cartan generated quantum circuits against exact unitary evolution and Suzuki–Trotter product evolution. We find that the Cartan approach resembles that of exact unitary evolution. For this, we calculate time evolution of spin expectation values and spin variances of the conjugate observables and it demonstrates that the Cartan circuits accurately reproduce both single spin observables, the collective fluctuation dynamics and dominant many-body correlations responsible for spin squeezing and entanglement generation. Comparisons based on error analysis reveal that optimized Cartan circuits can achieve practically vanishing errors than the fixed-order Suzuki Trotter decompositions while employing a substantially reduced sequence of unitary operations. The substantial difference between the two results with increasing ensemble size suggests that Cartan based circuit compilation becomes progressively more efficient in simulating the system with large ensembles.

A key advantage of the proposed framework is that the resulting generators corresponding to collective operations can be implemented in ensemble based quantum hardware. Neutral-atom arrays~\cite{bluvstein2024logical}, Rydberg atomic ensembles, cavity-mediated spin systems~\cite{PhysRevB.100.014415}, and other collectively controlled quantum platforms provide direct access to exchange interactions and nonlinear spin couplings~\cite{norcia2018cavity} that closely match the Cartan building blocks derived here. Consequently, QND measurements are available for the implementation of important gates in the construction of ensemble circuits. In conclusion, we find that Cartan decomposition not only offers an efficient numerical representation of many-body dynamics, but also provides a physically motivated compilation strategy for quantum simulation and quantum information processing in large atomic ensembles.

Looking forward, the present formalism opens several directions for future research. Extensions to higher order Cartan generators and enlarged dynamical Lie algebras for more than two ensembles may widen the applicability of the model and circuit compression for strongly interacting many-body systems. The framework can also be generalized to long-range interacting spin models~\cite{PhysRevA.77.062330}, dissipative spin system dynamics~\cite{chaudhary2026}, quantum information protocols~\cite{PhysRevA.111.042421,manish2021remote} and cavity-QED architectures~\cite{PhysRevA.111.053712}, where permutation symmetry remains an important organizing principle. More broadly, the combination of Lie-algebraic structure, collective-spin symmetries, and hardware-efficient circuit construction provides a promising pathway toward scalable quantum simulation protocols for next-generation neutral-atom and ensemble-based quantum processors.

\appendix

\section{Increasing Hamiltonian algebra for distinguishable spins}
\label{appendix_1}
Permutation symmetry substantially restricts the structure of the Hamiltonian algebra as expressed in~\eqref{permutation_liealgebra}.
In contrast, when permutation symmetry is absent, the particles are distinguishable and the operator space is spanned by independent Pauli strings
\begin{align}
     \sigma^\alpha_{1} \sigma^\alpha_{2}\sigma^\alpha_{3} \dots \sigma^\alpha_{N}, \alpha \in \{x,y,z\}
\end{align}
Consequently, we consider the Hamiltonian~\eqref{ham_ensemble} with distinct spin labels 
\begin{align}
    H_{\text{dis}} = \sum_{n,n'=1}^N (\sigma^x_{1,n} \sigma^x_{2,n'} + \sigma^y_{1,n} \sigma^y_{2,n'} ) 
\end{align}
There are $N^2$ distinct two-body couplings $\sigma^\alpha_{1,n} \sigma^\alpha_{2,n'}$ for each Pauli component and are independent operator degrees of freedom and can no longer be represented by a single collective tensor component as in the permutation symmetric case.
Under global $\mathfrak{su}(2)$ rotation for qubit space $\mathbb{C}^2$ with dimension $2^N$, the tensor decomposition therefore contains multiple copies of the same tensor rank such that the Hamiltonian Lie algebra is contained in
\begin{align}
   \mathfrak{g}_\text{dis}((\mathbb{C}^2)^{\otimes N}) = \underset{q}{\oplus} (\mathcal{M}_q \otimes \mathcal{T}^{(q)})
\end{align}
where $\mathcal{M}_q$ denotes the multiplicity of the rank-$q$ sector. The corresponding dynamical Lie algebra can contain a much larger number of independent operators with dimension $4^N-1$.

To understand it, we consider each system of three distinguishable spins $N=3$. For one spin, the Lie algebra is defined as
\begin{align}
    \mathfrak{g}_\text{dis}(\mathbb{C}^2) = \mathcal{T}^{(0)} \oplus \mathcal{T}^{(1)}
\end{align}
where $\mathcal{T}^{(0)}$ is generated by the identity and $\mathcal{T}^{(1)}$ by the three Pauli matrices. 
For three distinguishable spins,
\begin{align}
    \mathfrak{g}_\text{dis}((\mathbb{C}^2)^{\otimes 3})) & = (\mathcal{T}^{(0)} \oplus \mathcal{T}^{(1)})^{\otimes 3} \nonumber \\
    & = \mathcal{T}^{(0)} \oplus 3 \mathcal{T}^{(1)} \oplus 3(\mathcal{T}^{(1)} \otimes \mathcal{T}^{(1)}) \oplus (\mathcal{T}^{(1)})^{\otimes 3} \nonumber \\
    & =  4 \mathcal{T}^{(0)} \oplus 9 \mathcal{T}^{(1)} \oplus 5 \mathcal{T}^{(2)} \oplus \mathcal{T}^{(3)} 
\end{align}
with dimension $\sum_q m_q (2q+1)=64=4^3$ and $m_q$ is the multiplicity factor. After removing the identity operator, the traceless operator space has dimension $4^3-1$.

However, in the fully permutation-symmetric three spin subspace, the three spins form a single collective spin $J=N/2$, and the operator space contains only one copy of each tensor rank, i.e
\begin{align}
     \mathcal{T}^{(0)} \oplus  \mathcal{T}^{(1)} \oplus  \mathcal{T}^{(2)} \oplus \mathcal{T}^{(3)} 
\end{align}
with dimension $\sum_q (2q+1)=16=(N+1)^2$.
Thus, permutation symmetry eliminates the multiplicities of the irreducible tensor sectors and reduces the dimension of the operator space from $4^N$ to $(N+1)^2$.
%\begin{align}
 %   [S^\alpha_l,S^\beta_l ] \propto S^\gamma_l
 %    \label{eq:comm1}
%\end{align}
%where $\alpha,\beta \in \{x,y,z\}$ and $l$ represent the ensemble label.

%The other contribution comes from the commutation between linear and bilinear product as,
%\begin{align}
   % [S^\alpha_l,S^\beta_l S^\delta_m ] & =  [S^\alpha_l,S^\beta_l] S^\delta_m + S^\beta_l[S^\alpha_l, S^\delta_m] \nonumber \\
  %  & \propto S^\gamma_l S^\delta_m
  %   \label{eq:comm2}
%\end{align}

%Another contribution is developed from the commutation among the bilinear products giving rise to the higher order interaction terms,
%\begin{align}
  %  [S^\alpha_l S^\beta_m, S^\gamma_l S^\delta_m ] & =  [S^\alpha_l,S^\gamma_l] S^\beta_m S^\delta_m + S^\gamma_l S^\alpha_l[S^\beta_m, S^\delta_m ] \nonumber \\
    % & \propto S^\zeta_l (S^\beta_m S^\delta_m) + (S^\gamma_l S^\alpha_l) S^\eta_m \nonumber \\
   %  & \propto S^\zeta_l \{S^\beta_m, S^\delta_m\} + \{S^\gamma_l, S^\alpha_l \} S^\eta_m
  %   \label{eq:comm3}
%\end{align}
% One finds that this representation for the collective spin operators  makes the computation rather complex and hence the non vanishing anti commutator terms in Eq.~\eqref{eq:comm3} lead to analytically complex higher order terms in the spin Hamiltonian algebra that is not closed. 

%Since the collective spin %sector lies in permutational invariance space of dimension $(N+1)$ irrespective of the full Hilbert space $2^N$. 
%algebra follows $(S^\alpha_l)^2 \neq I $ in contrast to the qubit Pauli operators  $(\sigma^\alpha_{l,n})^2 = I$, this  

\section{Error bounds in Trotter and Cartan methods}
\label{appendix_2}
The Trotter error~\eqref{error} is governed by the higher commutators generated when the Hamiltonian is decomposed into noncommuting pieces.

For one trotter step of size, 
 \begin{align}
    \delta t = t/n_t 
 \end{align}
BCH expansion for trotterized unitary~\eqref{unitary_suzukitrotter} gives,
\begin{align}
  U_{\text{Trotter}}(t) & = (e^{-iH_x \delta t}e^{-iH_y \delta t})^{n_{t}} \nonumber \\
  & = \exp\left [-i(H_x + H_y)\delta t - \frac{\delta t ^2}{2}[H_x,H_y] + O(\delta t^3)\right]
\end{align}
where $H_x= S^x_1S^x_2$ and $H_y = S^y_1S^y_2$.

After $n_t$ steps, the expression~\eqref{error}
\begin{align}
   || U_{\text{Exact}}(t)- U_{\text{Trotter}}(t)|| = O \left(\frac{t^2}{n_t} ||[H_x,H_y]||\right)
\end{align}
Hence, the leading trotterized error~\eqref{error} is,
 \begin{align}
      \varepsilon_{\text{Trotter}}(t) \lesssim \frac{t^2}{n_t N}||[H_x,H_y]||
      \label{trottererrorbound}
 \end{align}
 up to higher-order corrections in $t/n_t$.
 
For the permutation symmetric $N$-atom ensemble,
\begin{align}
    ||S^x|| = ||S^y|| = ||S^z|| = N
\end{align}
and therefore, 
\begin{align}
    ||[H_x,H_y]|| & \leq \frac{1}{4} (||[S^x_1,S^y_1]||\times ||S^x_2|| \times ||S^y_2|| \nonumber \\
    & + ||S^y_1||\times||S^x_1|| \times||[S^x_2,S^y_2]|| )  \nonumber \\
    & =  O(N^3)
\end{align}
Thus at leading order, Trotter error~\eqref{trottererrorbound} becomes
%\begin{align}
   % \varepsilon_{\text{Trotter}}(t) = O\left(\frac{N^2 t^2}{n_t}\right)
%\end{align}
\begin{align}
    \varepsilon_{\text{Trotter}}(t)  \lesssim \frac{N^2 t^2}{2n_t}
\end{align}
 providing an upper bound.

On the other hand, Cartan method involves constructing~\eqref{cartandecom_ham} a Hamiltonian $\mathcal{H}^c_{\mathrm{ens}}$ defined in~\eqref{cartanhamiltonian} such that 
\begin{align}
    \Delta H = \mathcal{H}_{\mathrm{ens}}-\mathcal{H}^c_{\mathrm{ens}}
\end{align}
Using the Duhamel identity~\cite{umarov2012fractional}, one can write the difference in the exponential operators as
\begin{align}
    e^{-i\mathcal{H}_\text{ens}t}- e^{-i\mathcal{H}^c_\text{ens}t}  =  -i  
     \int_0^t e^{-i (t-t')\mathcal{H}_\text{ens}} \Delta H e^{-i t' \mathcal{H}^c_\text{ens}} dt'
\end{align}
Since the unitary operator does not change the operator norm and hence
\begin{align}
    ||e^{-i (t-t')\mathcal{H}_\text{ens}} \Delta H e^{-i t' \mathcal{H}^c_\text{ens}} || = ||\Delta H||
\end{align}
which expresses the Cartan error~\eqref{error} as
\begin{align}
     \varepsilon_{\text{Cartan}}(t)  \lesssim \frac{t}{N} ||\Delta H||
     \label{cartanerrorbound}
\end{align}
In terms of small errors in the optimized parameters~\eqref{kensemble},
\begin{align}
    \delta \eta_j = \eta_j - \eta^{\text{exact}}_j
    \label{parametererror}
\end{align}
and hence
\begin{align}
    \mathcal{H}^c_\text{ens} (\mathbf{\eta}) \simeq \mathcal{H}^c_\text{ens}(\mathbf{\eta^{\text{exact}}}) + \sum_j \frac{\partial \mathcal{H}^c_\text{ens}}{\partial \eta_j} \delta \eta_j
\end{align}
such that the exact Cartan representation satisfies
\begin{align}
    \mathcal{H}^c_\text{ens}(\mathbf{\eta^{\text{exact}}}) = \mathcal{H}_\text{ens}
\end{align}
Therefore, 
\begin{align}
    ||\Delta H || \simeq \sum_j \left|\left|\frac{\partial \mathcal{H}^c_\text{ens}}{\partial \eta_j}\right|\right| \delta \eta_j
\end{align}
Thus, Cartan error~\eqref{cartanerrorbound} can also be written in terms of the small errors in the optimized parameters~\eqref{parametererror} as
\begin{align}
    \varepsilon_{\text{Cartan}}(t)  \lesssim \frac{t}{N} \sum_j \left|\left|\frac{\partial \mathcal{H}^c_\text{ens}}{\partial \eta_j}\right|\right| \delta \eta_j
\end{align}
If the generator set is complete~\eqref{cartan_gen_ensemble} and the optimization finds the exact parameters, i.e. $\delta \eta_j = 0$ in Eq.~\eqref{parametererror}, then
\begin{align}
    \Delta H = 0
\end{align}
and 
\begin{align}
    \varepsilon_{\text{Cartan}}(t)  = 0
\end{align}

 \begin{acknowledgments}
M.C. acknowledges the funding support from the FWO and the F.R.S.-FNRS as part of the Excellence of Science program (EOS project 40007526) at University of Liège, Belgium during the initial calculations for this work.
\end{acknowledgments}

\bibliography{ref}

@article{RevModPhys.86.153,
  title = {Quantum simulation},
  author = {Georgescu, I. M. and Ashhab, S. and Nori, Franco},
  journal = {Rev. Mod. Phys.},
  volume = {86},
  issue = {1},
  pages = {153--185},
  numpages = {33},
  year = {2014},
  month = {Mar},
  publisher = {American Physical Society},
  doi = {10.1103/RevModPhys.86.153},
}

@article{feynman1982,
  title = {Simulating physics with computers},
  author = {Richard P. Feynman},
  journal = {International Journal of Theoretical Physics},
  year = {1982},
  volume = {21},
  number = {6},
  pages = {467--488},
  isbn = {1572-9575},
  doi = {10.1007/BF02650179},
}

@article{Preskill2018quantumcomputingin,
  title = {Quantum {C}omputing in the {NISQ} era and beyond},
  author = {Preskill, John},
  journal = {{Quantum}},
  issn = {2521-327X},
  publisher = {{Verein zur F{\"{o}}rderung des Open Access Publizierens in den Quantenwissenschaften}},
  volume = {2},
  pages = {79},
  month = aug,
  year = {2018},
  doi={10.22331/q-2018-08-06-79}
}

@article{lloyd1996universal,
  title={Universal quantum simulators},
  author={Lloyd, Seth},
  journal={Science},
  volume={273},
  number={5278},
  pages={1073--1078},
  year={1996},
  publisher={American Association for the Advancement of Science},
  doi={10.1126/science.273.5278.1073}
}

@article{PhysRevLett.129.070501,
  title = {Fixed Depth Hamiltonian Simulation via Cartan Decomposition},
  author = {K\"okc\"u, Efekan and Steckmann, Thomas and Wang, Yan and Freericks, J. K. and Dumitrescu, Eugene F. and Kemper, Alexander F.},
  journal = {Phys. Rev. Lett.},
  volume = {129},
  issue = {7},
  pages = {070501},
  numpages = {7},
  year = {2022},
  month = {Aug},
  publisher = {American Physical Society},
  doi = {10.1103/PhysRevLett.129.070501},
}

@article{l88j-p4f3,
  title = {Identification and optimization of accurate spin models for Fermi-Hubbard ladders using matrix product states},
  author = {Agirre, Andoni and Frederiksen, Thomas and Giedke, G\'eza and Gra\ss{}, Tobias},
  journal = {Phys. Rev. B},
  volume = {113},
  issue = {24},
  pages = {245422},
  numpages = {12},
  year = {2026},
  month = {Jun},
  publisher = {American Physical Society},
  doi = {10.1103/l88j-p4f3},
}

@book{carter1995lectures,
  title={Lectures on Lie groups and Lie algebras},
  author={Carter, Roger W and MacDonald, Ian G and Segal, Graeme B and Taylor, M},
  year={1995},
  publisher={Cambridge University Press},
  doi={10.1017/CBO9781139172882}
}

@article{PhysRevResearch.4.023054,
  title = {Engineering strong beamsplitter interaction between bosonic modes via quantum optimal control theory},
  author = {Basilewitsch, Daniel and Zhang, Yaxing and Girvin, S. M. and Koch, Christiane P.},
  journal = {Phys. Rev. Res.},
  volume = {4},
  issue = {2},
  pages = {023054},
  numpages = {17},
  year = {2022},
  month = {Apr},
  publisher = {American Physical Society},
  doi = {10.1103/PhysRevResearch.4.023054},
}

@article{bluvstein2024logical,
  title={Logical quantum processor based on reconfigurable atom arrays},
  author={Bluvstein, Dolev and Evered, Simon J and Geim, Alexandra A and Li, Sophie H and Zhou, Hengyun and Manovitz, Tom and Ebadi, Sepehr and Cain, Madelyn and Kalinowski, Marcin and Hangleiter, Dominik and others},
  journal={Nature},
  volume={626},
  number={7997},
  pages={58--65},
  year={2024},
  doi={10.1038/s41586-023-06927-3},
  publisher={Nature Publishing Group UK London}
}

@article{PhysRevB.100.014415,
  title = {Cavity-mediated dissipative spin-spin coupling},
  author = {Grigoryan, Vahram L. and Xia, Ke},
  journal = {Phys. Rev. B},
  volume = {100},
  issue = {1},
  pages = {014415},
  numpages = {7},
  year = {2019},
  month = {Jul},
  publisher = {American Physical Society},
  doi = {10.1103/PhysRevB.100.014415},
}

@article{drury2008constructive,
  title={Constructive quantum Shannon decomposition from Cartan involutions},
  author={Drury, Byron and Love, Peter},
  journal={Journal of Physics A: Mathematical and Theoretical},
  volume={41},
  number={39},
  pages={395305},
  year={2008},
  publisher={IOP Publishing},
  doi={10.1088/1751-8113/41/39/395305}
}

@article{norcia2018cavity,
  title={Cavity-mediated collective spin-exchange interactions in a strontium superradiant laser},
  author={Norcia, Matthew A and Lewis-Swan, Robert J and Cline, Julia RK and Zhu, Bihui and Rey, Ana M and Thompson, James K},
  journal={Science},
  volume={361},
  number={6399},
  pages={259--262},
  year={2018},
  doi={10.1126/science.aar3102},
  publisher={American Association for the Advancement of Science}
}

@article{khaneja2001cartan,
  title={Cartan decomposition of SU (2n) and control of spin systems},
  author={Khaneja, Navin and Glaser, Steffen J},
  journal={Chemical Physics},
  volume={267},
  number={1-3},
  pages={11--23},
  year={2001},
  publisher={Elsevier},
  doi={10.1016/S0301-0104(01)00318-4}
}

@article{PhysRevA.97.063613,
  title = {Spin-resolved single-atom imaging of $^{6}\mathrm{Li}$ in free space},
  author = {Bergschneider, Andrea and Klinkhamer, Vincent M. and Becher, Jan Hendrik and Klemt, Ralf and Z\"urn, Gerhard and Preiss, Philipp M. and Jochim, Selim},
  journal = {Phys. Rev. A},
  volume = {97},
  issue = {6},
  pages = {063613},
  numpages = {8},
  year = {2018},
  month = {Jun},
  publisher = {American Physical Society},
  doi = {10.1103/PhysRevA.97.063613},
}

@article{umarov2012fractional,
  title={On fractional Duhamelʼs principle and its applications},
  author={Umarov, Sabir},
  journal={Journal of Differential Equations},
  volume={252},
  number={10},
  pages={5217--5234},
  year={2012},
  publisher={Elsevier},
  doi={10.1016/j.jde.2012.01.029}
}

@article{daugli2008general,
  title={A general framework for recursive decompositions of unitary quantum evolutions},
  author={Da{\u{g}}l{\i}, Mehmet and D'Alessandro, Domenico and Smith, Jonathan DH},
  journal={Journal of Physics A: Mathematical and Theoretical},
  volume={41},
  number={15},
  pages={155302},
  year={2008},
  doi={10.1088/1751-8113/41/15/155302}
}

@article{alsheikh2025redcard,
  title={RedCarD: A Quantum Assisted Algorithm for Fixed-Depth Unitary Synthesis via Cartan Decomposition},
  author={Alsheikh, Omar and K{\"o}kc{\"u}, Efekan and Bakalov, Bojko N and Kemper, Alexander F},
  journal={arXiv preprint},
  year={2025},
doi={10.48550/arXiv.2505.04240}
}

@article{julsgaard2001experimental,
  title={Experimental long-lived entanglement of two macroscopic objects},
  author={Julsgaard, Brian and Kozhekin, Alexander and Polzik, Eugene S},
  journal={Nature},
  volume={413},
  number={6854},
  pages={400},
  year={2001},
doi={https://doi.org/10.1038/35096524
},
  publisher={Nature Publishing Group}
}

@article{PhysRevLett.112.233602,
  title = {Theory of Single-Shot Phase Contrast Imaging in Spinor Bose-Einstein Condensates},
  author = {Ilo-Okeke, Ebubechukwu O. and Byrnes, Tim},
  journal = {Phys. Rev. Lett.},
  volume = {112},
  issue = {23},
  pages = {233602},
  numpages = {5},
  year = {2014},
  month = {Jun},
  publisher = {American Physical Society},
  doi = {10.1103/PhysRevLett.112.233602},
}

@article{kuzmich1998atomic,
  title={Atomic quantum non-demolition measurements and squeezing},
  author={Kuzmich, A and Bigelow, NP and Mandel, L},
  journal={EPL (Europhysics Letters)},
  volume={42},
  number={5},
  pages={481--486},
  year={1998},
  doi={10.1209/epl/i1998-00277-9}
}

@article{pezze2018quantum,
  title={Quantum metrology with nonclassical states of atomic ensembles},
  author={Pezz{\`e}, Luca and Smerzi, Augusto and Oberthaler, Markus K and Schmied, Roman and Treutlein, Philipp},
  journal={Rev. Mod. Phys.},
  volume={90},
  number={3},
  pages={035005},
  year={2018},
doi={https://doi.org/10.1103/RevModPhys.90.035005},
  publisher={APS}
}

@article{gao2023optical,
  title={Optical and atomic decoherence in quantum nondemolition measurement induced atomic ensemble entanglement},
  author={Gao, Shuai and Li, Shuang and Chaudhary, Manish and Prest, Matthew and Ilo-Okeke, Ebubechukwu O and Ivannikov, Valentin and Byrnes, Tim},
  journal={AVS Quantum Science},
  volume={5},
  number={3},
  year={2023},
  doi={10.1116/5.0147830}, 
  publisher={AIP Publishing},
}

@article{PhysRevLett.122.163601,
  title = {Observation of Atom Number Fluctuations in a Bose-Einstein Condensate},
  author = {Kristensen, M. A. and Christensen, M. B. and Gajdacz, M. and Iglicki, M. and Paw\l{}owski, K. and Klempt, C. and Sherson, J. F. and Rz{a}\ifmmode \dot{z}\else \.{z}\fi{}ewski, K. and Hilliard, A. J. and Arlt, J. J.},
  journal = {Phys. Rev. Lett.},
  volume = {122},
  issue = {16},
  pages = {163601},
  numpages = {6},
  year = {2019},
  month = {Apr},
  publisher = {American Physical Society},
  doi = {10.1103/PhysRevLett.122.163601},
}

@article{PhysRevA.77.062330,
  title = {Dynamics of entanglement in a one-dimensional Ising chain},
  author = {Furman, G. B. and Meerovich, V. M. and Sokolovsky, V. L.},
  journal = {Phys. Rev. A},
  volume = {77},
  issue = {6},
  pages = {062330},
  numpages = {6},
  year = {2008},
  month = {Jun},
  publisher = {American Physical Society},
  doi = {10.1103/PhysRevA.77.062330},
}

@article{gao2025photon,
  title={Photon loss effects on light-mediated non-Gaussian entangled Bose--Einstein condensates projecting with different counts},
  author={Gao, Shuai and Chaudhary, Manish and Pyrkov, Alexey N and Ilo-Okeke, Ebubechukwu O and Meng, Xin and Feng, Jingyan and Khan, Muhammad Jamil and Byrnes, Tim and Lou, Chaogang},
  journal={Physica Scripta},
  volume={100},
  number={6},
  pages={065111},
  year={2025},
  publisher={IOP Publishing},
  doi={10.1088/1402-4896/add393}
}

@article{PhysRevA.111.053712,
  title = {Cavity quantum electrodynamics with atom arrays in free space},
  author = {Castells-Graells, David and Cirac, J. Ignacio and Wild, Dominik S.},
  journal = {Phys. Rev. A},
  volume = {111},
  issue = {5},
  pages = {053712},
  numpages = {20},
  year = {2025},
  month = {May},
  publisher = {American Physical Society},
  doi = {10.1103/PhysRevA.111.053712},
}

@article{PhysRevA.111.042421,
  title = {Macroscopic quantum teleportation with ensembles of qubits},
  author = {Chaudhary, Manish and Lin, Zhiyuan and Li, Shuang and Zhang, Mohan and Mao, Yuping and Ivannikov, Valentin and Byrnes, Tim},
  journal = {Phys. Rev. A},
  volume = {111},
  issue = {4},
  pages = {042421},
  numpages = {17},
  year = {2025},
  month = {Apr},
  publisher = {American Physical Society},
  doi = {10.1103/PhysRevA.111.042421},
}

@article{chaudhary2026,
      title={Dissipative State Engineering of Complex Entanglement with Markovian Dynamics}, 
      author={Manish Chaudhary},
      year={2026},
      archivePrefix={arXiv},
      primaryClass={quant-ph},
      journal={arxiv-preprint arXiv.2601.09597},
      doi={10.48550/arXiv.2601.09597},
}

@article{kitzinger2020,
  title={Two-axis two-spin squeezed states},
  author={Kitzinger, Jonas and Chaudhary, Manish and Kondappan, Manikandan and Ivannikov, Valentin and Byrnes, Tim},
  journal={Phys. Rev. Research},
  volume={2},
  number={3},
  pages={033504},
  year={2020},
doi={https://doi.org/10.1103/PhysRevResearch.2.033504},
  publisher={APS}
}

@book{timquantumoptics2020,
  title={Quantum atom optics: Theory and applications to quantum technology},
  author={Byrnes, Tim and Ilo-Okeke, Ebubechukwu O},
  year={2021},
doi={https://doi.org/10.1017/9781108975353},
  publisher={Cambridge university press}
}

@article{PhysRevLett.98.070201,
  title = {Classical Simulation of Infinite-Size Quantum Lattice Systems in One Spatial Dimension},
  author = {Vidal, G.},
  journal = {Phys. Rev. Lett.},
  volume = {98},
  issue = {7},
  pages = {070201},
  numpages = {4},
  year = {2007},
  month = {Feb},
  publisher = {American Physical Society},
  doi = {10.1103/PhysRevLett.98.070201},
}

@article{PRXQuantum.2.010342,
  title = {Real- and Imaginary-Time Evolution with Compressed Quantum Circuits},
  author = {Lin, Sheng-Hsuan and Dilip, Rohit and Green, Andrew G. and Smith, Adam and Pollmann, Frank},
  journal = {PRX Quantum},
  volume = {2},
  issue = {1},
  pages = {010342},
  numpages = {15},
  year = {2021},
  month = {Mar},
  publisher = {American Physical Society},
  doi = {10.1103/PRXQuantum.2.010342},
}

@article{mcardle2019variational,
  title={Variational ansatz-based quantum simulation of imaginary time evolution},
  author={McArdle, Sam and Jones, Tyson and Endo, Suguru and Li, Ying and Benjamin, Simon C and Yuan, Xiao},
  journal={npj Quantum Information},
  volume={5},
  number={1},
  pages={75},
  year={2019},
  publisher={Nature Publishing Group UK London},
  doi={10.1038/s41534-019-0187-2}
}

@book{thompson2008angular,
  title={Angular momentum: an illustrated guide to rotational symmetries for physical systems},
  author={Thompson, William J},
  year={2008},
  publisher={John Wiley \& Sons},
  doi={10.1002/9783527617821}
}

@article{PhysRevLett.131.110602,
  title = {Measurement-Based Deterministic Imaginary Time Evolution},
  author = {Mao, Yuping and Chaudhary, Manish and Kondappan, Manikandan and Shi, Junheng and Ilo-Okeke, Ebubechukwu O. and Ivannikov, Valentin and Byrnes, Tim},
  journal = {Phys. Rev. Lett.},
  volume = {131},
  issue = {11},
  pages = {110602},
  numpages = {6},
  year = {2023},
  month = {Sep},
  publisher = {American Physical Society},
  doi = {10.1103/PhysRevLett.131.110602},
}

@article{cirstoiu2020variational,
  title={Variational fast forwarding for quantum simulation beyond the coherence time},
  author={Cirstoiu, Cristina and Holmes, Zoe and Iosue, Joseph and Cincio, Lukasz and Coles, Patrick J and Sornborger, Andrew},
  journal={npj Quantum Information},
  volume={6},
  number={1},
  pages={82},
  year={2020},
  publisher={Nature Publishing Group UK London}
}

@article{PhysRevA.108.052607,
  title = {Near-optimal quantum circuit construction via Cartan decomposition},
  author = {Mansky, Maximilian Balthasar and Castillo, Santiago Londo\~no and Puigvert, Victor Ramos and Linnhoff-Popien, Claudia},
  journal = {Phys. Rev. A},
  volume = {108},
  issue = {5},
  pages = {052607},
  numpages = {9},
  year = {2023},
  month = {Nov},
  publisher = {American Physical Society},
  doi = {10.1103/PhysRevA.108.052607},
}

@article{manish2021remote,
  title={Remote state preparation of two-component Bose-Einstein condensates},
  author={Chaudhary, Manish and Fadel, Matteo and Ilo-Okeke, Ebubechukwu O and Pyrkov, Alexey N and Ivannikov, Valentin and Byrnes, Tim},
  journal={Phys. Rev. A},
  volume={103},
  number={6},
  pages={062417},
  year={2021},
doi={https://doi.org/10.1103/PhysRevA.103.062417},
  publisher={APS}
}

@article{trotter1959product,
  title={On the product of semi-groups of operators},
  author={Trotter, Hale F},
  journal={Proceedings of the American Mathematical Society},
  volume={10},
  number={4},
  pages={545--551},
  year={1959},
  publisher={JSTOR},
  doi={10.1090/s0002-9939-1959-0108732-6}
}

@article{PhysRevX.11.011020,
  title = {Theory of Trotter Error with Commutator Scaling},
  author = {Childs, Andrew M. and Su, Yuan and Tran, Minh C. and Wiebe, Nathan and Zhu, Shuchen},
  journal = {Phys. Rev. X},
  volume = {11},
  issue = {1},
  pages = {011020},
  numpages = {49},
  year = {2021},
  month = {Feb},
  publisher = {American Physical Society},
  doi = {10.1103/PhysRevX.11.011020},
}

@article{dhand2014stability,
  title={Stability of the Trotter--Suzuki decomposition},
  author={Dhand, Ish and Sanders, Barry C},
  journal={Journal of Physics A: Mathematical and Theoretical},
  volume={47},
  number={26},
  pages={265206},
  year={2014},
  publisher={IOP Publishing},
  doi={10.1088/1751-8113/47/26/265206}
}

@article{suzuki1985decomposition,
  title={Decomposition formulas of exponential operators and Lie exponentials with some applications to quantum mechanics and statistical physics},
  author={Suzuki, Masuo},
  journal={Journal of mathematical physics},
  volume={26},
  number={4},
  pages={601--612},
  year={1985},
  publisher={AIP Publishing},
  doi={10.1063/1.526596}
}

@article{ostmeyer2023optimised,
  title={Optimised Trotter decompositions for classical and quantum computing},
  author={Ostmeyer, Johann},
  journal={Journal of Physics A: Mathematical and Theoretical},
  volume={56},
  number={28},
  pages={285303},
  year={2023},
  publisher={IOP Publishing},
  doi={10.1088/1751-8121/acde7a}
}

@article{PhysRevLett.111.253001,
  title = {Accurate Atom Counting in Mesoscopic Ensembles},
  author = {Hume, D. B. and Stroescu, I. and Joos, M. and Muessel, W. and Strobel, H. and Oberthaler, M. K.},
  journal = {Phys. Rev. Lett.},
  volume = {111},
  issue = {25},
  pages = {253001},
  numpages = {5},
  year = {2013},
  month = {Dec},
doi={https://doi.org/10.1103/PhysRevLett.111.253001},
  publisher = {American Physical Society},
}

@article{lane2025ordering,
  title={Ordering Matters: Structure, Accuracy and Gate Cost in Second-Order Suzuki Product Formulas},
  author={Lane, Matthew A and Browne, Dan E},
  year={2025},
  journal= {arXiv preprint},
  doi={10.48550/arXiv.2505.04240}
}

@article{chen2026trotter,
  title={Trotter error and gate complexity of the SYK and sparse SYK models},
  author={Chen, Yiyuan and Helsen, Jonas and Ozols, Maris},
  journal={Quantum},
  volume={10},
  pages={1999},
  year={2026},
  publisher={Verein zur F{\"o}rderung des Open Access Publizierens in den Quantenwissenschaften},
  doi={10.22331/q-2026-02-09-1999}
}

@article{PhysRevLett.114.090502,
  title = {Simulating Hamiltonian Dynamics with a Truncated Taylor Series},
  author = {Berry, Dominic W. and Childs, Andrew M. and Cleve, Richard and Kothari, Robin and Somma, Rolando D.},
  journal = {Phys. Rev. Lett.},
  volume = {114},
  issue = {9},
  pages = {090502},
  numpages = {5},
  year = {2015},
  month = {Mar},
  publisher = {American Physical Society},
  doi = {10.1103/PhysRevLett.114.090502},
}

@article{kalev2021quantum,
  title={Quantum algorithm for simulating hamiltonian dynamics with an off-diagonal series expansion},
  author={Kalev, Amir and Hen, Itay},
  journal={Quantum},
  volume={5},
  pages={426},
  year={2021},
  publisher={Verein zur F{\"o}rderung des Open Access Publizierens in den Quantenwissenschaften},
  doi={10.22331/q-2021-04-08-426}
}

@article{PhysRevA.105.032420,
  title = {Algebraic compression of quantum circuits for Hamiltonian evolution},
  author = {K\"okc\"u, Efekan and Camps, Daan and Bassman Oftelie, Lindsay and Freericks, J. K. and de Jong, Wibe A. and Van Beeumen, Roel and Kemper, Alexander F.},
  journal = {Phys. Rev. A},
  volume = {105},
  issue = {3},
  pages = {032420},
  numpages = {14},
  year = {2022},
  month = {Mar},
  publisher = {American Physical Society},
  doi = {10.1103/PhysRevA.105.032420},
}

@article{d2006decompositions,
  title={Decompositions of unitary evolutions and entanglement dynamics of bipartite quantum systems},
  author={D’Alessandro, Domenico and Romano, Raffaele},
  journal={Journal of mathematical physics},
  volume={47},
  number={8},
  year={2006},
  publisher={AIP Publishing},
  doi={10.1063/1.2245205}
}

@article{earp2005constructive,
  title={A constructive algorithm for the Cartan decomposition of SU (2N)},
  author={Earp, Henrique N S{\'a} and Pachos, Jiannis K},
  journal={Journal of mathematical physics},
  volume={46},
  number={8},
  year={2005},
  publisher={AIP Publishing},
  doi={10.1063/1.2008210}
}

@article{PhysRevA.78.052101,
  title = {Collective processes of an ensemble of spin-$1∕2$ particles},
  author = {Chase, Bradley A. and Geremia, J. M.},
  journal = {Phys. Rev. A},
  volume = {78},
  issue = {5},
  pages = {052101},
  numpages = {14},
  year = {2008},
  month = {Nov},
  publisher = {American Physical Society},
  doi = {10.1103/PhysRevA.78.052101},
}

@article{moroder2012permutationally,
  title={Permutationally invariant state reconstruction},
  author={Moroder, Tobias and Hyllus, Philipp and T{\'o}th, G{\'e}za and Schwemmer, Christian and Niggebaum, Alexander and Gaile, Stefanie and G{\"u}hne, Otfried and Weinfurter, Harald},
  journal={New Journal of Physics},
  volume={14},
  number={10},
  pages={105001},
  year={2012},
  publisher={IOP Publishing},
  doi={10.1088/1367-2630/14/10/105001}
}

@article{PhysRevA.88.012305,
  title = {Genuine multiparticle entanglement of permutationally invariant states},
  author = {Novo, Leonardo and Moroder, Tobias and G\"uhne, Otfried},
  journal = {Phys. Rev. A},
  volume = {88},
  issue = {1},
  pages = {012305},
  numpages = {8},
  year = {2013},
  month = {Jul},
  publisher = {American Physical Society},
  doi = {10.1103/PhysRevA.88.012305},
}

@article{PhysRevA.105.022443,
  title = {Stroboscopic quantum nondemolition measurements for enhanced entanglement generation between atomic ensembles},
  author = {Chaudhary, Manish and Mao, Yuping and Kondappan, Manikandan and Paz, Amiel S. P. and Ivannikov, Valentin and Byrnes, Tim},
  journal = {Phys. Rev. A},
  volume = {105},
  issue = {2},
  pages = {022443},
  numpages = {17},
  year = {2022},
  month = {Feb},
  publisher = {American Physical Society},
  doi = {10.1103/PhysRevA.105.022443},
}

@article{PhysRevA.108.032420,
  title = {Macroscopic maximally-entangled-state preparation between two atomic ensembles},
  author = {Chaudhary, Manish and Ilo-Okeke, Ebubechukwu O. and Ivannikov, Valentin and Byrnes, Tim},
  journal = {Phys. Rev. A},
  volume = {108},
  issue = {3},
  pages = {032420},
  numpages = {13},
  year = {2023},
  month = {Sep},
  publisher = {American Physical Society},
  doi = {10.1103/PhysRevA.108.032420},
}

@article{aristizabal2021quantum,
  title={Quantum nondemolition measurement based generation of entangled states in two Bose Einstein condensates},
  author={Aristizabal-Zuluaga, Juan E and Skobleva, Iuliia and Richter, Lars and Ji, Yangxu and Mao, Yuping and Kondappan, Manikandan and Ivannikov, Valentin and Byrnes, Tim},
  journal={Journal of Physics B: Atomic, Molecular and Optical Physics},
  volume={54},
  number={10},
  pages={105502},
  year={2021},
  publisher={IOP Publishing},
  doi={10.1088/1361-6455/abf6b5}
}

\end{document}